\documentclass[final]{svjour3}
\usepackage{graphicx}
\usepackage{rotating}
\usepackage{amssymb}
\usepackage{mathptmx}
\usepackage{color}
\usepackage{comment}
\usepackage[numbers]{natbib}
\usepackage[normalem]{ulem}

\makeatletter
\journalname{Journal of Low Temperature Physics}

\usepackage[usenames,dvipsnames,svgnames,table]{xcolor}	

\bibpunct{}{}{,}{s}{}{,}

\def\Journal#1#2#3#4{{#1} {\bf #2}, #3 (#4)}

\def\NIMA{{\em Nucl. Instrum. Methods} A}

\def\PRL{\em Phys. Rev. Lett.}

\def\AHEP{\em Adv. High Energy Phys.} 

\def\JLT{\em J. Low Temp. Phys.}
\def\Cry{\em Cryogenics}

\def\be{\begin{equation}}
\def\ee{\end{equation}}
\def\bea{\begin{eqnarray}}
\def\eea{\end{eqnarray}}

\begin{document}

\newcommand{\hdblarrow}{H\makebox[0.9ex][l]{$\downdownarrows$}-}
\title{The CUORE cryostat}


\author{A.~D'Addabbo$^{1,2,*}$, C.~Alduino$^{3}$, A.~Bersani$^{4}$, M.~Biassoni$^{5}$, C.~Bucci$^{1}$, A.~Caminata$^{4}$, L.~Canonica$^{6,7,1}$, L.~Cappelli$^{1,8,9}$, G.~Ceruti$^{5}$, N.~Chott$^{3}$, S.~Copello$^{1,2}$, O.~Cremonesi$^{5}$, J.~S.~Cushman$^{10}$, D.~D'Aguanno$^{1,11}$, C.~J.~Davis$^{10}$, S.~Dell'Oro$^{12}$, S.~Di~Domizio$^{13,4}$, A.~Drobizhev$^{8,9}$, M.~Faverzani$^{14,5}$, E.~Ferri$^{14,5}$, M.~A.~Franceschi$^{15}$, L.~Gladstone$^{6}$, P.~Gorla$^{1}$, C.~Ligi$^{15}$, L.~Marini$^{8,9,1}$, T.~Napolitano$^{15}$, A.~Nucciotti$^{14,5}$, I.~Nutini$^{1,2}$, J.~L.~Ouellet$^{6}$, C.~E.~Pagliarone$^{1,11}$, L.~Pattavina$^{16,2,1}$, C.~Rusconi$^{3,1}$, D.~Santone$^{1,17}$, B.~Schmidt$^{9}$, V.~Singh$^{8}$, D.~Speller$^{10}$, L.~Taffarello$^{18}$, F.~Terranova$^{14,9}$, J.~Wallig$^{9}$, B.~Welliver$^{9}$, T.~Wise$^{10,19}$} 

\institute{$^{*}$ \email{antonio.daddabbo@lngs.infn.it}}

\authorrunning
\titlerunning
\maketitle

\noindent
$^{1}$ INFN -- Laboratori Nazionali del Gran Sasso, Assergi (L'Aquila) I-67100, Italy \\
$^{2}$ Gran Sasso Science Institute, L'Aquila I-67100, Italy \\
$^{3}$ Department of Physics and Astronomy, University of South Carolina, Columbia, SC 29208, USA \\
$^{4}$ INFN -- Sezione di Genova, Genova I-16146, Italy \\
$^{5}$ INFN -- Sezione di Milano Bicocca, Milano I-20126, Italy \\
$^{6}$ Massachusetts Institute of Technology, Cambridge, MA 02139, USA \\
$^{7}$ Max-Planck-Institut f\"ur Physik, D-80805, M\"unchen, Germany \\
$^{8}$ Department of Physics, University of California, Berkeley, CA 94720, USA \\
$^{9}$ Nuclear Science Division, Lawrence Berkeley National Laboratory, Berkeley, CA 94720, USA \\
$^{10}$ Wright Laboratory, Department of Physics, Yale University, New Haven, CT 06520, USA \\
$^{11}$ Dipartimento di Ingegneria Civile e Meccanica, Universit\`{a} degli Studi di Cassino e del Lazio Meridionale, Cassino I-03043, Italy \\
$^{12}$ Center for Neutrino Physics, Virginia Polytechnic Institute and State University, Blacksburg, Virginia 24061, USA \\
$^{13}$ Dipartimento di Fisica, Universit\`{a} di Genova, Genova I-16146, Italy \\
$^{14}$ Dipartimento di Fisica, Universit\`{a} di Milano-Bicocca, Milano I-20126, Italy \\
$^{15}$ INFN -- Laboratori Nazionali di Frascati, Frascati (Roma) I-00044, Italy \\
$^{16}$ Physik Department, Technische Universit\"at M\"unchen, D85748 Garching, Germany \\
$^{17}$ Dipartimento di Scienze Fisiche e Chimiche, Universit\`{a} dell'Aquila, L'Aquila I-67100, Italy \\
$^{18}$ INFN -- Sezione di Padova, Padova I-35131, Italy \\
$^{19}$ Department of Physics, University of Wisconsin, Madison, WI 53706, USA \\

\vspace{6mm}

\begin{abstract}

The Cryogenic Underground Observatory for Rare Events (CUORE) is a bolometric experiment for neutrinoless double-beta decay in $^{130}$Te search, currently taking data at the underground facility of Laboratori Nazionali del Gran Sasso (LNGS). The CUORE cryostat successfully cooled down a mass of about 1 ton at $\sim$7\,mK, delivering an uniform and constant base temperature. This result marks a fundamental milestone in low temperature detectors techniques, opening the path for future ton-scale bolometric experiments searching for rare events. In this paper we present the CUORE cryogenic infrastructure, briefly describing its critical subsystems.


\end{abstract}
\vspace{4mm}

\section{Introduction}

CUORE is an experiment devoted to search for neutrinoless double-beta decay and other rare events \cite{da, detector}. CUORE operates a ton-scale array of 988 TeO$_2$ bolometers at $\sim$10\,mK providing exceptionally low background and low vibration conditions.  In order to meet this unprecedented challenge, the CUORE detector array is cooled down by a multistage cryogen-free cryostat unique of its kind. Due to the large mass, a custom made precooling system brings the cryostat to a temperature of 35\,K using liquid He (LHe) vapours. Later the cryostat is cooled down to 4\,K by five Pulse Tubes (PTs). Finally the base temperature is delivered by a custom designed continuous-cycle $^3$He/$^4$He dilution refrigerator. Strict material selection and cleaning procedures are applied to all the structures facing the detector in order to preserve the low radioactivity environment necessary to reach CUORE goals. More than seven tons of lead shielding, placed at low temperature, protect the inner cubic meter size experimental volume from residual radioactive contamination from the cryostat. Special suspensions mechanically decouple the detector from the cryostat. Residual vibration-induced thermal noise on the bolometers is minimized by means of a dedicated software that optimizes the working frequencies and phases of the active PTs. The cooling power ($>$ 3\,$\mu$W at 10\,mK), the total mass ($\sim$17\,tons) and the experimental volume ($\geq$1\,m$^3$) make the CUORE cryostat the biggest and most powerful dilution cryostat in the world.

\section{The cryostat}

\begin{figure}[h!]
\begin{center}
\includegraphics[width=0.8\linewidth]{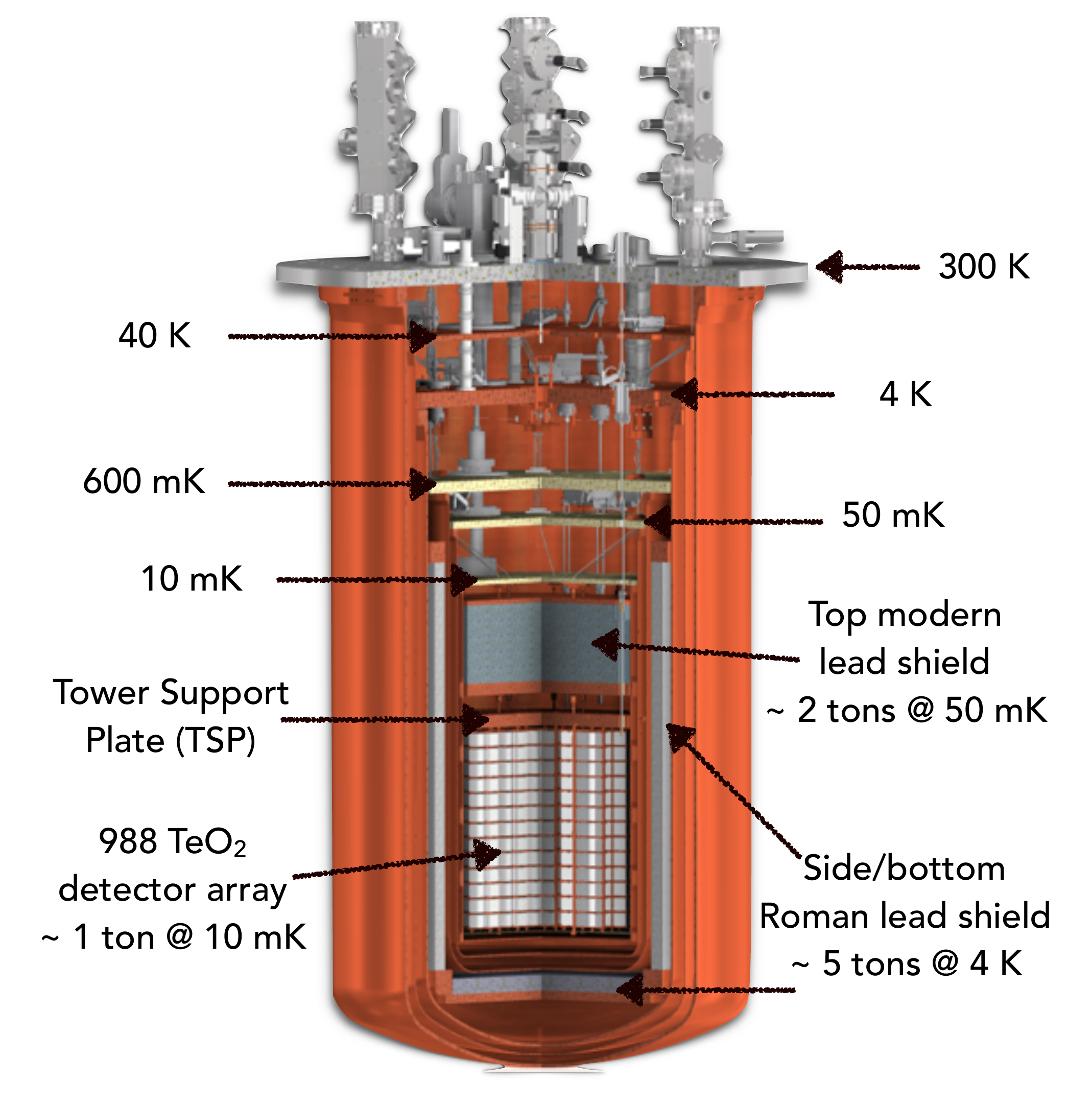}
\end{center}
\caption{The CUORE cryostat structure with all its vessels and lead shieldings.}
\label{fig:Cryostat}
\end{figure}

The CUORE cryostat (see Fig. \ref{fig:Cryostat}) is composed by 6 coaxial vessels, kept at 300\,K, 40\,K, 4\,K, 600\,mK (Still), 50\,mK (Exchanger, HEX), 10\,mK (Mixing Chamber, MC), respectively. 300\,K and 4\,K shields are vacuum tight, creating two separated vacuum chambers: the outermost one is referred to as Outer Vacuum Chamber (OVC), with a volume of 5.9\,m$^3$ and a minimum pressure lower than 10$^{-6}$\,mbar; the innermost one is referred to as Inner Vacuum Chamber (IVC), with a volume of 3.4\,m$^3$ and a minimum pressure lower than 10$^{-8}$\,mbar. 

The CUORE cryostat construction and assembly represented an unprecedented technological challenge. It is really difficult to determine which subsystems need major effort. We can cite just few among the most challenging and time consuming ones, like the Roman lead casting of the ingots for the inner lead shielding, the fabrication and installation of the Minus-K suspension system, the Dilution Refrigerator custom made by Leiden Cryogenics, and last but not least, the design of the cryostat vessels and plates, including the evaluations of the number and positioning of the inlet/outlet apertures and feed troughs. 

The cryostat fabrication involved several highly qualified manufacturers \cite{constr} and required the application of strict radio-purity material selection criteria \cite{criteria}. The 40 K, the 4 K, the Still and the HEX vessels has been built using Oxygen-Free Copper (OFC) \cite{da}, (OFHC compound C10100) is the highest purity grade of copper at 99.99\%.
Given its high conductivity at low temperatures (RRR  $\geq$ 400) and for the low hydrogen content, the highly purified ETP1 copper alloy (Electronic Tough Pitch, also called NOSV Cu by Aurubis) \cite{mc, C0det} has been chosen for the most critical inner parts facing the detector, such as the MC flange and plate, the Tower Support Plate, the final Detector Suspension part and the detector frames.

\subsection{The detector suspension}

The detector is not anchored to the cryostat. It is attached to the so called Tower Support Plate (TSP), which is is located in the IVC, 76.8 mm below the Top lead shielding, inside 10\,mK vessel. The TSP is mechanically suspended from an external structure, called Y-beam, by means of three Detector Suspension (DS) bars. Each DS bar consists of a segmented structure which is thermalized at each thermal stage of the cryostat. From the Y-beam down to the Still stage, a single DS bar structure is made by stainless steel (SS 316LN) bars with a diameter varying from 12\,mm to 6\,mm. In order to minimize the heat load, the TSP is then suspended from the last thermal contact at the Still stage by a couple of Kevlar K49 ropes per each DS bar (three couples in total) with a diameter of 6\,mm. The last segment of each DS bar is made of high purity copper rod, chosen for the low radioactivity content. The TSP thermalization to 10\,mK stage is provided by four copper connectors to the 10\,mK vessel, each one consisting in a few tens of thin NOSV Cu sheets in order to provide mechanical isolation. 

The DS share the OVC vacuum thanks to two vacuum tight sealings. Thanks to a tube with bellows at the two ends, the OVC vacuum extends out of the cryostat up to the bottom part of the Y-beam, where the bellow ends into a blank flange rigidly connected to the Y-beam. On the other end, the bellow is vacuum tighten to the 300K plate using an o-ring. The second sealing, internal to the cryostat, separates the OVC from the IVC vacuum by means of an indium seal at the level of the 4K plate, which extends the OVC vacuum into the IVC, below the 4 K plate, by mean of a blank bellow.

The Y-beam is positioned at room temperature, upon the cryostat support structure, called Main Support Plate (MSP), and anchored to three mechanical insulators by Minus-K Technology. The Minus-K insulators consist of a particular arrangement of elastic constants that acts like a soft spring
and results in a low-pass filter for the vibrations with an effective cut-off frequency close to 0.5 Hz. Preliminary measurements in the (0-25) Hz frequency range with an impulse excitation to the system, have shown a $\sim$ 30 dB attenuation of the TSP acceleration with respect to the MSP, as measured in the vertical axis \cite{susp}.

\subsection{Cold lead shieldings}

To suppress $\gamma$ background coming from the natural radioactive decay chain of $^{238}$U and $^{232}$Th present as residual contaminants in the cryostat, two different lead shields are placed at cryogenic temperatures.

The lateral and bottom shielding is provided by the so-called Roman lead shield. It is made of $^{210}$Pb-depleted ancient lead ingots (I sec b.C.) from a Roman shipwreck found close to Sardinia coast. Custom ingots were cast in a nitrogen atmosphere and the shield was assembled in late 2015. The shield is 6\,cm in thickness, with a radius and an height of about 64\,cm and 178\,cm, respectively. Its weight is around 5\,tons, it is placed inside the IVC, suspended from the 600\,mK plate and thermalized at 4\,K.
An additional cold lead shield is placed inside the IVC. It is a 30\,cm thick modern lead, thermalized at 50\,mK and protecting the upper part of the detector from the cryostat and external radioactivity. With a radius of about 36\,cm, it accounts for more than 2\,tons of lead.

\section{The multistage refrigerator}

The CUORE cyostat is a cryogen-free infrastructure able to guarantee a base temperature lower than $\sim$10\,mK for a ton-scale detector and a 5\,years live-time. It is capable to maintain 0.98\,ton and 7.4\,tons respectively at 40\,K and at 4\,K, and cool down approximately 4.5\,tons of material below 50\,mK. The total mass cooled down below 10\,mK is around 1520 kg of material, including the MC plate and vessel, the TSP, the TeO${_2}$ crystals and their frames. 

Several square meters of superinsulation aluminised Si multilayer foils have been used on 40\,K and 4\,K vessels and plates to limit their irradiation on lower temperature stages.
The wiring consists in a system of woven ribbon cables used to bring the signal from the MC to the outside of the cryostat. The ribbon cables consists in 13 twisted pairs of NbTi wires (100\,$\mu$m diameter) with a CuNi coating (5\,$\mu$m thickness). 
With a total of 100 ribbon cables, the wiring accounts for a total of 2600 wires. The wiring thermalization is provided by custom made copper connectors at each temperature stage interface.

Given the cryostat demand in terms of thermal load, a cooling power exceeding 96 W at 45 K, 3.6 W at 4.2 K, 3 mW at the Still stage, 125 $\mu$W at the HEX stage and 4 $\mu$W at the MC stage is needed. To address these requirements, the CUORE cryostat rely on three different customized cryorefrigerator systems which are able to reach different temperature stages. In the following we will briefly describe them.

\subsection{Fast Cooling System}

The first stage is provided by the so-called Fast Cooling System (FCS) \cite{FCS}. It is a
pre cooling apparatus which forces the circulation of $^4$He gas into the IVC to pre-cool its mass down to $\sim$50\,K. The FCS uses the gas evaporating from a LHe tank , which is firstly filtered and then cooled down using three Gifford-McMahon cold heads in a custom made external cooling circuit. The He gas is then injected into the IVC of the cryostat by means of proper combinations of dedicated polytetrafluoroethylene tubes. The He gas acts as heat exchanger between the different stages of the IVC, thermalizing them to 4\,K stage. The He gas is constantly recycled through an outlet tube to force the circulation of fresh cooled gas inside the IVC.
The FCS operates the cryostat down to $\sim$50\,K in approximately two weeks, allowing to keep the cool down time well below one month. At that point, the Pulse Tubes carry forward the cooling down.

\subsection{Pulse Tubes}

The second stage at $\sim$4\,K is obtained by means of five two-stages Cryomech PT415-RM Pulse Tubes (PTs), with nominal cooling power of 1.2\,W at 4.2\,K and 32\,W at 45\,K each. They come into play after the first week of cooling down with the FCS. In order to limit the vibrations injected in the detector, a series of mechanically decouplings have been applied. Before to reach the cryostat top plate, the PT flexlines enter a sand box filled with non-hygroscopic quartz powder. All the PT flexlines and the rotating valves are suspended from the ceiling in such a way to avoid any contact with the cryostat. Moreover, on its 300 K plate, the PT rotating valve are operated in a custom remote motor-head configuration with swan-neck outlet. All the PT flanges at room temperature are mechanically decoupled from the 300 K plate of the cryostat by mean of special polyurethane collars. Further decoupling is guarantied at 40\,K and 4\,K stages by using soft copper braid thermalisation.

\subsection{Dilution unit}

The base temperature is provided by a DRS-CF3000 continuous-cycle $^3He/^4He$ Dilution Refrigerator (DR), customized by Leiden Cryogenics for CUORE purposes. Its nominal cooling power is 3\,$\mu$W at 12\,mK (2\,mW at 120\,mK), and the minimum reachable temperature for the CUORE cryostat at full load is around 7\,mK (for more details refer to \cite{Ligi}). Operating the DU with no additional thermal load on the Still, the mixture flow is about 1 mmol/sec and the effective measured cooling powers correspond to 2\,mW at 100\,mK (3\,mW at 123\,mK) and 4\,$\mu$W at 10\,mK, larger than the requested ones. The latter value could be increased up to 6\,$\mu$W at 10\,mK dissipating 10\,mW on the Still, with a correspondent mixture flow just above 1500 mmol/sec \cite{Dello}.

\begin{figure}[h!]
\begin{center}
\includegraphics[width=0.8\linewidth]{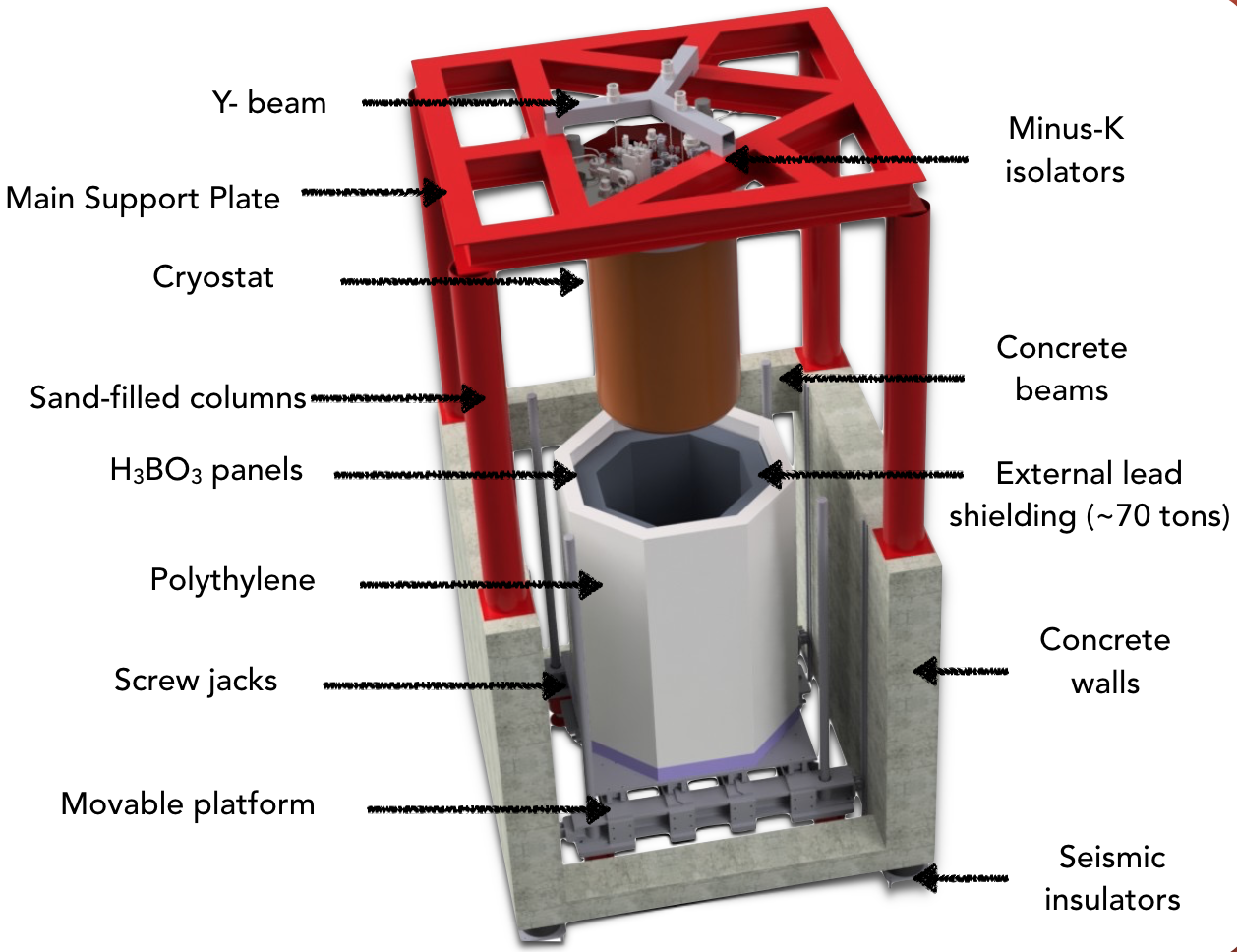}
\end{center}
\caption{Schematic of the CUORE support infrastructure.}
\label{fig:Cryostat2}
\end{figure}

\section{External infrastructure and shielding}

At room temperature, the cryostat is suspended to a structure called Main Support Plate (MSP). It can be identified by the red structure in Fig. \ref{fig:Cryostat2}. A Y-shaped beam, isolated from the MSP through the special Minus-K suspensions, supports the detector. In this way, vibrations transmission from the cryostat to the TeO$_2$ crystals are minimized. The whole MSP stands on a concrete structure by means of four sand-filled columns. The concrete structure is in turn insulated from the CUORE building using four seismic elastomers.

To further shield the detector for the $\gamma s$ sources external to the cryostat, an external lead shield ($\sim$70\,tons, thickness = 25\,cm) can be raised at room temperature by means of four screw jacks. From outer inwards, the lead is covered by a polythylene shielding and boric acid 
panels for neutron thermalization and capture, respectively.
The entire external shielding structure is placed on a movable platform, which allows its removal if needed.

\section{Other subsystems}

\subsection{Detector Calibration System}

Twelve kevlar strings, each one equipped with 25 thoriated ($^{232}$Th) tungsten capsules, can be deployed inside the cryostat to provide the detector with an energy calibration. The so-called Detector Calibration System (DCS) parameters, such as the sources activities, have been optimized by GEANT4 MC simulations. To avoid detector self-shielding, the strings can be deployed among the towers (6 internal, 3.5\,Bq), facing directly the detector. Additional 6 external strings with higher activity (19.4\,Bq) can be deployed out of the 50\,mK vessel, providing the calibration for the outer towers. More details on the DCS can be found in \cite{jc}.

\subsection{PT noise suppression}

The PT cryorefrigerators are one of the main source of vibrations, which are transmitted to the detector through the cryostat. The suppression of the residual amount of vibration-induced thermal noise on the bolometers is critical for optimising the detector performances. For this purpose, an active system drives the PT rotating valves with a low-noise microstepping motor, optimizing their relative operational phases in order to achieve coherent vibration addition. This technique, called active PT noise cancellation \cite{NelloPT}, has been applied to the first CUORE data taking campaign, which results are published in \cite{CUORE}.

\begin{figure}[h!]
\begin{center}$
\begin{array}{c}  
\includegraphics[width=0.48\linewidth]{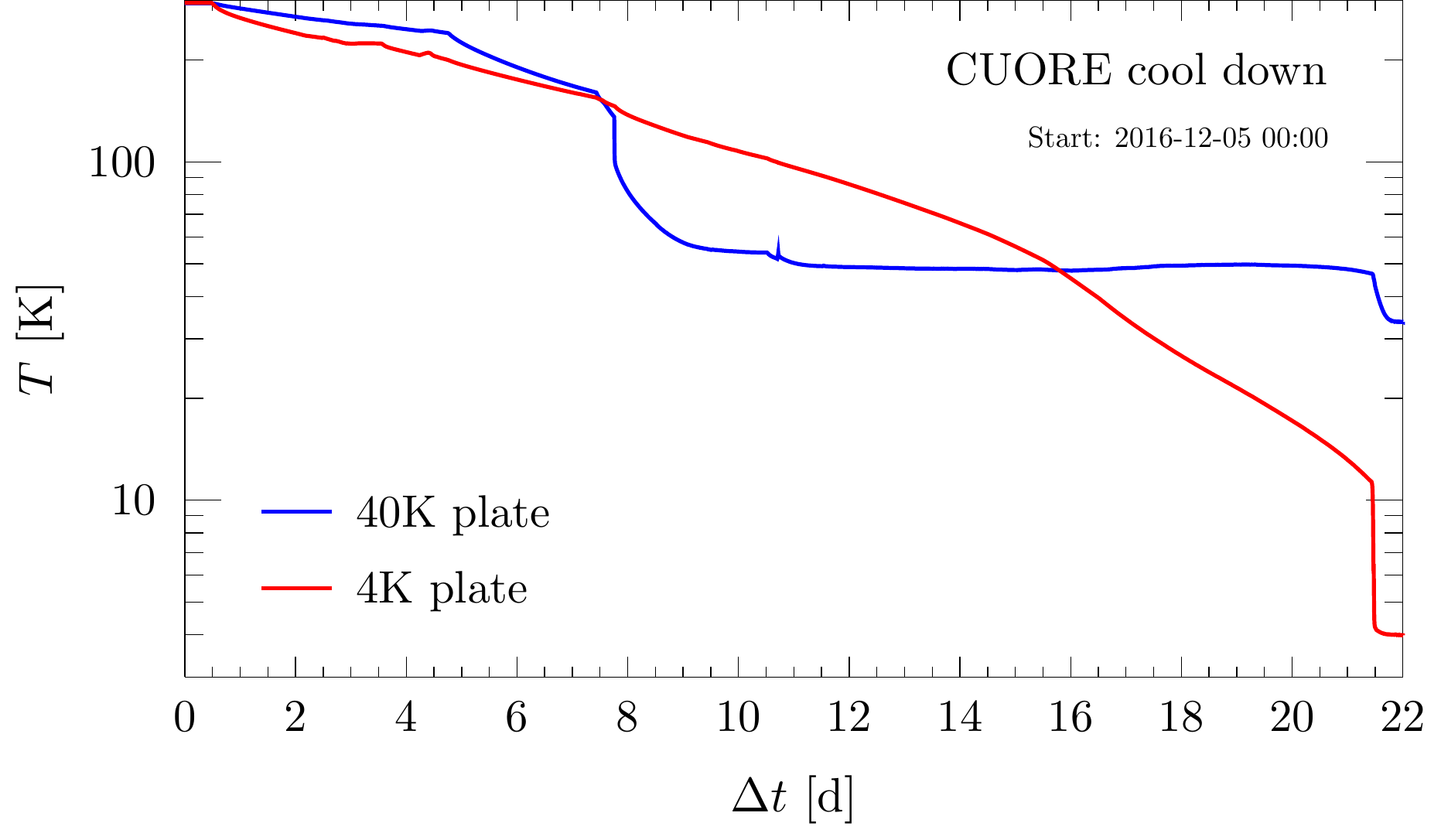}
\quad
\includegraphics[width=0.51\linewidth]{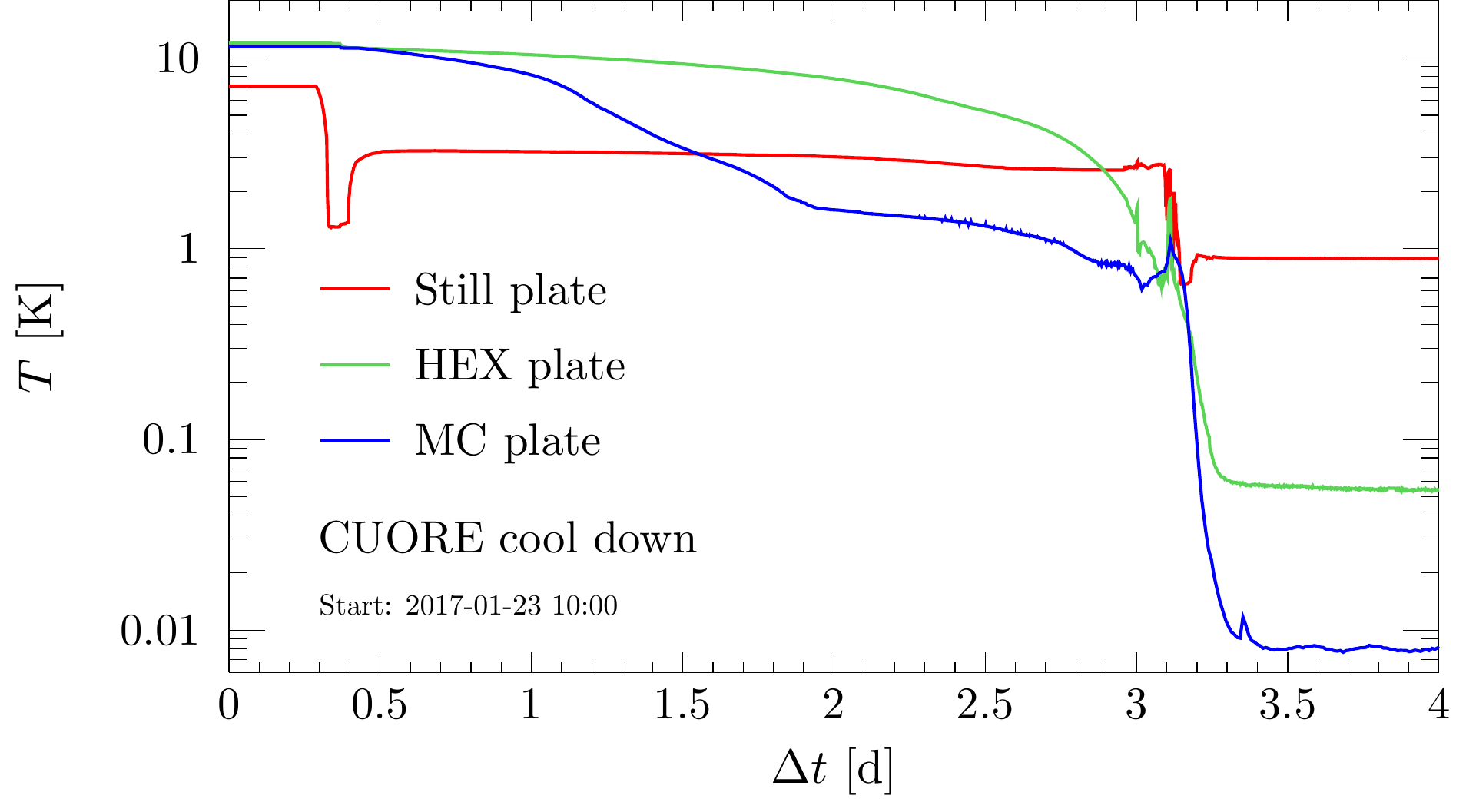}
\end{array}$
\end{center}
\caption{Temperature plots of the CUORE cooldown}
\label{fig:cooldown}
\end{figure}

\section{CUORE cooldown}

After the completion of the detector towers installation and the cryostat closure, the first phase of the official CUORE cool down started. The left plot in Fig. \ref{fig:cooldown} shows the temperatures of the 40\,K and 4\,K plates. This phase started turning on the FCS on December 5$^{th}$ 2016. After about 4 days, also the PTs were switched on. After the first week, the FCS was turned off and we continued the cool down using the PTs only. We reached 3.4\,K after approximately 22\,days after the beginning of the cool down. 

A period of about one months was then dedicated to the pumping of residual exchange gas in the IVC and to the optimization of several subsystems. On January 23$^{rd}$ 2017, we started the DU, which took another 4 days to reach the base temperature of $\sim$6.9\,mK (see Fig. \ref{fig:cooldown}, right plot). The total net cool down time was approximately 26\,days, during which almost 10$^9$\,J of enthalpy have been removed from the cryostat.

\section*{Acknowledgments}

The CUORE Collaboration thanks the directors and staff of the Laboratori Nazionali del Gran Sasso and the technical staff of our laboratories. This work was supported by the Istituto Nazionale di Fisica Nucleare (INFN); the National Science Foundation; the Alfred P. Sloan Foundation; the University of Wisconsin Foundation; and Yale University. This material is also based upon work supported by the US Department of Energy (DOE) Office of Science; and by the DOE Office of Science, Office of Nuclear Physics.  This research used resources of the National Energy Research Scientific Computing Center (NERSC).


\end{document}